\documentclass[showpacs,showkeys,amsmath,amssymb,10pt,aps]{revtex4}

\usepackage[T1]{fontenc}
\usepackage[cp1250]{inputenc}
\usepackage{amsfonts}
\usepackage{amsbsy}
\usepackage{amssymb}
\usepackage{amsmath}
\usepackage{bbm}
\usepackage{bm}
\usepackage{epstopdf}
\usepackage{tikz}

\begin{document}

\title[From a posteriori to a priori solutions for a two-level system interacting with a single-photon wavepacket]{From a posteriori to a priori solutions for a two-level system interacting with a single-photon wavepacket}

\author{Anita Magdalena D\k{a}browska}

\affiliation{Institute of Theoretical Physics and Astrophysics, University of Gda\'nsk, ul. Wita Stwosza 57, 80-308 Gda\'nsk, Poland}

\email{anita.dabrowska@ug.edu.pl}

\begin{abstract}
	 We present the analytical formulas for the conditional and unconditional states of a two-level atom interacting with a single-photon wavepacket. We express the {\it a priori} state of the system by means of the quantum trajectories related to the process of detection of photons in the output field. We give the formulas for the mean number of photons detected up to the given time and we derive the expressions for the mean time of detection of the photons.
\end{abstract}

	\maketitle

	\section{Background and motivation}
	
	Together with a development of experimental methods of generating and manipulating of the wavepackets of definite numbers of photons \cite{RNB05, CWSS13, YMM13, PSZ13, RR15, LMS15, OOM16, LMS17, SKL18}, many theoretical descriptions of their interactions with quantum systems were proposed. The scattering process of $N$ photons on a quantum system was described in the pure-state wavefunction approaches \cite{KB16,NKM15} and diagrammatic approaches \cite{SS09,SNA17,RS16}. Generalized master equations \cite{GEPZ98,DHR02,GJN11,WMSS11,WMHS12,BCBC12,RB17} and stochastic master equations \cite{GJNC12a,GJN12b,CHJ12,GJN13,GJN14,GZ15,DZA15,SZX16,PDZ16,BC17,DSC17,D18,DSC19,D19,Z19,DZA19} were used to study the excitation of the two-level atom interacting with a $N$ photon packets.  
	
	In this paper we treat a problem of an interaction of a quantum system with an environment prepared in a single-photon state \cite{L00,M08}. As it was shown in \cite{GEPZ98}, the reduced dynamics of an open system is given then by the set of coupled first-ordered linear differential equations. In the paper we derive the solution to these hierarchy equations for the two-level atom by applying the solutions to the filtering equations obtained in \cite{DSC17,DSC19}. Let us remind that the quantum filtering theory \cite{Bel89, Bel90, BB91, Car93, BP02, GZ10, WM10}, formulated in the framework of quantum stochastic It\^{o} calculus \cite{HP84,Par92}, provides a description for the conditional evolution of a quantum system interacting with the Bose field. In this approach, usually applied for the Markovian evolution, we deal with a division of the external field into the input and output parts \cite{GarCol85, Bar06}. The input part is interpreted as the field before the interaction with the system and the output part is the field after this interaction. By performing the measurements on the output field we gain the information about the system of our interest. The conditional evolution of the system depending on the results of the continuous in time measurements of the output field is given by the filtering equation (stochastic master equation). The form of this equation depends on the type of the measurement and on the initial state of the Bose field. The solutions to this equation are called quantum trajectories or the {\it a posteriori} states. An average of the quantum trajectories taken over all possible realizations of the stochastic process related to the considered measurement gives the {\it a priori} state of the system, which for the Markovian case fulfills the Gorini-Kossakowski-Sudarshan-Lindbland master equation \cite{GKS76, Lin76}. The standard methods of quantum filtering theory are tailored to the case when the input field is in a Gaussian state (coherent, thermal or squeezed) and they can not be directly used to case when the field  is prepared in the single-photon state. As shown in \cite{DSC17,DSC19,D19} to determine the stochastic evolution of the system interacting with the field in a non-classical state one can apply the model of quantum repeated interactions and measurements \cite{AP06, P08, PP09, P10} (called also the collision model \cite{KLS16,C17,ACMZ17,FPMZ17,GCMC18,CC19}). The collision model allows not only to determine the stochastic evolution of the system but it also enables us to find a physical intuitive interpretation to the quantum trajectories and stochastic evolution. The evolution of a quantum system interacting with the field prepared in the single-photon state is non-Markovian due to the temporal field correlations. The impact of these correlations on the evolution of open systems was analyzed in the framework of the collision model, for instance, in \cite{FPMZ17,RFZB12,BCMS14,BCMS17,MV17}. Discussions about the discrete filtering equations and their continuous limits one can find, for instance, in \cite{P08,P10,GCMC18,B02,GS04,BHJ09}. The time discretization procedure leading to the collision model in quantum optics and its connection with the input-output formalism and quantum trajectories are given in \cite{C17,GCMC18,FPMZ17}.

	In this paper we apply the formulas for the quantum trajectories derived in \cite{DSC17,DSC19} to describe the conditional and unconditional dynamics of the two-level atom, and to specify the statistics of the output photons. In  \cite{DSC17} the authors derived the filtering equations for the system driven by the field in a single-photon state starting from a discrete in time description of evolution of the system interacting with an environment modeled by a chain of qubits prepared initially in an entangled state being a discrete analogue of continuous mode single-photon state. The evolution of the composed system was given by repeated unitary interactions between the system and the bath qubits interrupted by the measurements performed on the bath qubits. The authors derived the set of recursive stochastic equations describing the conditional evolution of the system and then determined their differential versions. The system and the form of the coupling operator were not specified in \cite{DSC17,DSC19} and their results can be applied to quantum systems such as atoms, ions or cavity modes. Unlike the papers \cite{GJNC12a,CHJ12,BCBC12,BC17} we do present numerical simulations to the filtering equations.  	
	
	The paper is organized as follows.  In Sec. 2 we recall some of the results of \cite{DSC17,DSC19} necessary in understanding the main part of this paper. We briefly describe the model of repeated interactions and measurements, we write down the set of filtering equations for the counting stochastic process, the set of master equations, and give the formula for the {\it a priori} state. In Sec. 3 we write the explicit analytical expressions for the {\it a posteriori} and  the {\it a priori} states for the two-level atom interacting with a single-photon packet. We make a comment about their physical meaning and write the formula for the mean time of detection of the output photons. Moreover, we provide the formula for positive-operator valued measure (POVM) associated with the counting process. In the Appendix we analyze POVM for the detection of the output photons indexed by the number of counts.

	\section{Filtering model}

\subsection{Repeated interactions and measurements}	
	
We consider a quantum system $\mathcal{S}$ coupled to the environment modeled by a chain of $N$ qubits. From the mathematical point of view, we use ``toy Fock space'' \cite{M95,GS04,BHJ09}. We assume that the environment qubits do not interact between themselves, but they interact with the system $\mathcal{S}$, one after the other, each during a time interval of the length $\tau=T/N$. The Hilbert space of the composed system, consisting of the environment and $\mathcal{S}$, is thus $\mathcal{H}_{\mathcal{E}}\otimes \mathcal{H}_{\mathcal{S}}$,
where $\mathcal{H}_{\mathcal{S}}$ is the Hilbert space associated with ${\mathcal{S}}$ and
$\mathcal{H}_{\mathcal{E}}=\bigotimes_{k=0}^{N-1}\mathcal{H}_{\mathcal{E},k}$,
where $\mathcal{H}_{\mathcal{E},k}=\mathbb{C}^2$ is the Hilbert space of the $k$-th qubit interacting with $\mathcal{S}$ in the interval $[k\tau, (k+1)\tau)$.
The evolution of the composed system is given by the sequence of unitary operators, for $1\leq j\leq N-1$
\begin{equation}
U_{j\tau} = {V}_{j-1} {V}_{j-2} \ldots {V}_{0},\;\;\;\;\;U_{0}=\mathbbm{1},
\end{equation}
where 
\begin{equation}
{V}_{k}=\exp\left(-i \tau H_{k}\right),
\end{equation}
\begin{eqnarray}\label{hamiltonian}
H_{k} &=& \mathbbm{1}_{k}\otimes H_{\mathcal{S}} +  \frac{i}{\sqrt{\tau}}\left(\sigma_{k}^{+}\otimes L-
\sigma_{k}^{-}\otimes L^{\dagger}\right),
\end{eqnarray}
where $H_\mathcal{S}$ is the Hamiltonian associated with  $\mathcal{S}$, $L$ is a bounded operator acting on $\mathcal{H}_{\mathcal{S}}$, $\sigma^{-}_{k}=|0\rangle_{k}\langle 1|$, $\sigma^{+}_{k}=|1\rangle_{k}\langle 0|$, where $|0\rangle_{k}$ and $|1\rangle_{k}$ stand respectively for a ground and excited state of the $k$-th two-level system of the bath. The derivation the Hamiltonian (\ref{hamiltonian}) together with discussion about physical assumptions leading to it one can find, for instance, in \cite{C17,GCMC18,FTVRS18}. The interaction between the system and the field is described in the Weisskopf-Wigner approximation. We set here the Planck constant $\hbar=1$ and we omit the identity operators from $\bigotimes_{i\neq k}\mathcal{H}_{\mathcal{E},i}$ to simplify the notation.

We assume that the composed system is initially prepared in a factorized state of the form
\begin{equation}
|1_{\xi}\rangle\otimes|\psi_{0}\rangle,
\end{equation}
where $|\psi_{0}\rangle$ denotes the state of the system $\mathcal{S}$ and $|1_{\xi}\rangle$ is the state of the environment defined as
\begin{equation}
|1_{\xi}\rangle=  \sqrt{\tau}  \sum_{k=0}^{N-1}\xi_{k}|1_k \rangle,
\end{equation}
where
\begin{equation}
|1_k \rangle=|0\rangle_{0}\otimes\ldots |0\rangle_{k-1}\otimes|1\rangle_{k}\otimes |0\rangle_{k+1}\otimes \ldots|0\rangle_{N-1},
\end{equation}
and $\displaystyle{\sum_{k=0}^{N-1}}|\xi_{k}|^2\tau = 1$. Note that $|\xi_{k}|^2\tau$ is the probability of detecting a photon in the interval $[k\tau, (k+1)\tau)$. Here we deal with a discrete version  of a single photon state defined in \cite{GJN11,GJNC12a,GJN12b,GJN13,DZA15,Z19} for the symmetric Fock space $\mathcal{F}\left(L^{2}(\mathbb{R}_{+})\right)$.    

 We gain the information about the system $\mathcal{S}$ in indirect way, namely by performing a sequence of measurements on the bath qubits just after their interactions with $\mathcal{S}$.  
 The conditional state of $\mathcal{S}$ and the part of the environment chain which has not interacted with $\mathcal{S}$ up to time $j\tau $  is at the moment $j\tau $ given by \cite{DSC17}
\begin{equation}\label{posteriori2}
|\tilde{\Psi}_{j| \pmb{\eta}_j}\rangle = \frac{|\Psi_{j| \pmb{ \eta}_j}\rangle}{\sqrt{\langle\Psi_{j| \pmb{\eta}_j}|\Psi_{j| \pmb{\eta}_j}\rangle}},
\end{equation}
where $ |\Psi_{j|\pmb{\eta}_j} \rangle $ is
the unnormalized conditional vector from the space $\displaystyle{\bigotimes_{k=j}^{N-1}}\mathcal{H}_{\mathcal{E},k}\otimes \mathcal{H}_{\mathcal{S}}$ having the form
\begin{eqnarray}\label{posteriori3}
|\Psi_{j|\pmb{\eta}_j} \rangle =  \sqrt{\tau}  \sum_{k=j}^{N-1}\xi_{k}|1_k \rangle_{[j,N-1]} \otimes |\alpha_{j| \pmb{\eta}_j}\rangle+ | vac \rangle_{[j,N-1]} \otimes |\beta_{j| \pmb{\eta}_j}\rangle,
\end{eqnarray}
where
\begin{equation}
|1_k \rangle_{[j,N-1]}=|0\rangle_{j}\otimes\ldots |0\rangle_{k-1}\otimes|1\rangle_{k}\otimes |0\rangle_{k+1}\otimes \ldots|0\rangle_{N-1},
\end{equation}
and $|vac\rangle_{[j,N-1]}=|0\rangle_{j}\otimes |0\rangle_{j+1} \otimes |0\rangle_{j+2} \otimes \ldots|0\rangle_{N-1}$.
In this paper we are interested in the quantum trajectories related to the results of the measurement of the environment observable
\begin{equation}\label{observable}
\sigma_{k}^{+}\sigma_{k}^{-}=|1\rangle_{k}\langle 1|,\;\;k=0, 1, 2, \ldots.
\end{equation} 
By $\pmb{\eta}_j$ we denoted a stochastic $j$-vector $\pmb{\eta}_j = (\eta_j,\eta_{j-1},\ldots,\eta_1)$ representing results of all measurements of the observable (\ref{observable}) up to $j\tau$. The conditional vectors $|\alpha_{j| \pmb{\eta}_j}\rangle$,  $|\beta_{j| \pmb{\eta}_j}\rangle$ from $\mathcal{H}_{\mathcal{S}}$ can be obtained by two recursive equations. The form of the conditional vector (\ref{posteriori3}) indicates that $\mathcal{S}$ becomes entangled with the input field. Note that the vector $|\alpha_{j|\pmb{\eta}_j}\rangle$ is associated with the scenario that $\mathcal{S}$ has not met the bath qubit prepared in the upper state up to $j\tau$ and $|\beta_{j|\pmb{\eta}_j}\rangle$ is the vector related to the scenario that the system has already met the bath qubit in the upper state. This is the crucial point in our future discussion about interpretation of the quantum trajectories. Of course, by taking the partial trace of $|\tilde{\Psi}_{j|\pmb{\eta}_j} \rangle\langle \tilde{\Psi}_{j|\pmb{\eta}_j}|$ with respect to the environment, we obtain the {\it a posteriori} state of $\mathcal{S}$.  

\subsection{Filtering equations and quantum trajectories}
From the set of recursive equations for the conditional vectors $|\alpha_{j|\pmb{\eta}_j}\rangle$, $|\beta_{j|\pmb{\eta}_j}\rangle$, one can derive the set of discrete filtering equations for $\mathcal{S}$. In the continuous time limit, when $\tau \to 0$ and $N\to \infty$ such that $T=N\tau$ is fixed, we obtain the set of stochastic differential equations of the form \cite{DSC17}
	\begin{eqnarray}\label{filcont1}
	d\tilde{\rho}_{t}&=&\mathcal{L}\tilde{\rho}_{t}dt+[\tilde{\rho}^{01}_{t},L^{\dagger}]\xi_{t}dt
	+[L, \tilde{\rho}^{10}_{t}]\xi^{\ast}_{t}dt\nonumber\\
	&+& \bigg\{\frac{1}{k_{t}}\left(
	L\tilde{\rho}_{t}L^{\dagger}+L\tilde{\rho}^{10}_{t}\xi^{\ast}_{t}+\tilde{\rho}^{01}_{t}L^{\dagger}\xi_{t}
	+\tilde{\rho}^{00}_{t}|\xi_{t}|^2\right)
	-\tilde{\rho}_{t}\bigg\}\left(dn(t)-k_{t}dt\right),
	\end{eqnarray}
	
	\begin{eqnarray}\label{filcont2}
	d\tilde{\rho}^{01}_{t}&=& \mathcal{L}\tilde{\rho}^{01}_{t}dt+\left[L,\tilde{\rho}^{00}_{t}\right]\xi^{\ast}_{t}dt\\
	&+& \left\{\frac{1}{k_{t}}\left(L\tilde{\rho}^{01}_{t}L^{\dagger}+
	L\tilde{\rho}^{00}_{t}\xi^{\ast}_{t}\right)-\tilde{\rho}^{01}_{t}\right\}\left(dn(t)-k_{t}dt\right)\,\nonumber
	\end{eqnarray}
	\begin{equation}\label{filcont3}
	d\tilde{\rho}^{00}_{t}=\mathcal{L}\tilde{\rho}_{t}^{00}dt + \left(\frac{1}{k_{t}}L\tilde{\rho}^{00}_{t}L^{\dagger}-
	\tilde{\rho}^{00}_{t}\right)\left(dn(t)-k_{t}dt\right)
	\end{equation}
with the superoperator $\mathcal{L}$ acting as
\begin{equation}
\mathcal{L}\tilde{\rho}=-i[H_{S},\tilde{\rho}]-
\frac{1}{2}\left\{L^{\dagger}L,\tilde{\rho}\right\}
+L\tilde{\rho}L^{\dagger}
\end{equation}
and
\begin{equation}
k_{t}=\mathrm{Tr}\left(L^{\dagger}L\tilde{\rho}_{t}
+L\tilde{\rho}^{10}_{t}\xi_{t}^{\ast}+\tilde{\rho}^{01}_{t}L^{\dagger}\xi_{t}+\tilde{\rho}^{00}_{t}|\xi_{t}|^2\right).
\end{equation}
The set of coupled nonlinear stochastic differential equations determines the conditional dynamics of $\mathcal{S}$. The {\it a posteriori} normalized state of $\mathcal{S}$ at the moment $t$ is given by $\tilde{\rho}_{t}$. 
The operators $\tilde{\rho}^{10}_{t}$ and $\tilde{\rho}^{01}_{t}$ are related by $\tilde{\rho}^{10}_{t}=(\tilde{\rho}^{01}_{t})^{\dagger}$. Initially, $\tilde{\rho}_{0}=\tilde{\rho}^{00}_{0}=|\psi_{0}\rangle\langle\psi_{0}|$ and $\tilde{\rho}^{01}_{0}= \tilde{\rho}^{10}_{0}=0$. By $n(t)$ we denoted the stochastic counting process with the {\it a posteriori} mean value given by $\mathbbm{E}[dn(t)|\tilde{\rho}_{t}]=k_{t}dt$. Note that $n(t)$ is a regular process thus  $(dn(t))^2=dn(t)$, which means that at most one photon can be detected in the interval of the length $dt$. The probability of detecting more than two photons in the interval $[t,t+dt)$ in the output channel for is proportional to $(dt)^2$ and it is negligible \cite{DSC19}. 

Finally, passing to $T\to \infty$, we get for the amplitude $\xi_{t}$ the normalization condition
\begin{equation}
\int_{0}^{+\infty}|\xi_{t}|^2dt=1.
\end{equation} 
  
The solution to (\ref{filcont1}) can be written in the form 
\begin{equation}
\tilde{\rho}_{t|cond}=\frac{\rho_{t|cond}}{\mathrm{Tr}\rho_{t|cond}},
\end{equation}
where
\begin{equation}\label{posteriori4}
{\rho}_{t|cond}=|\alpha_{t|cond}\rangle\langle\alpha_{t|cond}|\int_{t}^{+\infty}ds|\xi_{s}|^2+|\beta_{t|cond}\rangle\langle\beta_{t|cond}|
\end{equation}
and $|\alpha_{t| cond}\rangle$, $|\beta_{t|cond}\rangle$ are the conditional vectors depending on all results of the continuous in time measurements of the output field up to $t$. The formulas for $|\alpha_{t|cond}\rangle$ and $|\beta_{t|cond}\rangle$ are rather involved but they can be written with making use of some simple diagrammatic representation (see details in \cite{DSC17,DSC19}).

\subsection{The a priori state and the set of master equations}
By taking the mean value of the {\it a posteriori} state $\tilde{\rho}_{t}$ over all possible outcomes (all possible realizations of the stochastic process related to the measured observable) we obtain the {\it a priori} state of $\mathcal{S}$. The {\it a priori} state, $\tilde{\varrho}_{t}$, in the representation of the counting stochastic process $n(t)$ has the form   
\begin{equation}\label{apriori}
\tilde{\varrho}_{t}=\rho_{t|0}+\sum_{m=1}^{+\infty}\int_{0}^{t}dt_{m}\ldots\int_{0}^{t_{3}}dt_{2}
\int_{0}^{t_{2}}dt_{1}\rho_{t|t_{m},\ldots,t_{2},t_{1}},
\end{equation}
where $\rho_{t|t_{m},\ldots,t_{2},t_{1}}$ are the conditional operators defined by (\ref{posteriori4}). In (\ref{apriori}) we have a sum over all possible pathways of the photon detection for the number of photons ranging from $m=0$ to $m=\infty$ from the time $0$ to $t$. The quantity $\mathrm{Tr}\rho_{t|0}$ is the probability of lack of any detections from $0$ to $t$, while $\mathrm{Tr}\rho_{t|t_{m},\ldots,t_{2},t_{1}}$ for all $m\geq 1$ is the probability density of detecting photons at times $t_{1}, t_{2}, \ldots, t_{m}$  (strictly at intervals $[t_{1},t_{1}+dt)$, $[t_{2},t_{2}+dt)$, $\ldots$, $[t_{m},t_{m}+dt)$) such that $0<t_{1}<t_{2}<\ldots<t_{m}$ and no other photons from $0$ to $t$. Clearly, for the system with a finite number degrees freedom, the sum in (\ref{apriori}) has limited number of non-zero terms. The set of master equations describing the {\it a priori} evolution of $\mathcal{S}$ has the form \cite{DSC17}
\begin{equation}\label{p1}
\dot{\tilde{\varrho}}_{t}=\mathcal{L}\tilde{\varrho}_{t}+
[\tilde{\varrho}^{01}_{t},L^{\dagger}]\xi_{t}
+[L, \tilde{\varrho}^{10}_{t}]\xi^{\ast}_{t},
\end{equation}
\begin{equation}\label{p2}
\dot{\tilde{\varrho}}^{01}_{t}= \mathcal{L}\tilde{\varrho}^{01}_{t}+
\left[L,\tilde{\varrho}^{00}_{t}\right]\xi^{\ast}_{t},
\end{equation}
\begin{equation}\label{p3}
\dot{\tilde{\varrho}}^{00}_{t}= \mathcal{L}\tilde{\varrho}_{t}^{00},
\end{equation}
where $\tilde{\varrho}_{t}$ is the {\it a priori} state of $\mathcal{S}$, $\tilde{\varrho}^{10}_{t}=(\tilde{\varrho}^{01}_{t})^{\dagger}$, and initially $\tilde{\varrho}_{0}=\tilde{\varrho}^{00}_{0}=|\psi_{0}\rangle\langle\psi_{0}|$, $\tilde{\varrho}^{01}_{0}= \tilde{\varrho}^{10}_{0}=0$. 
The general solution to this set can be written as 
\begin{equation}\label{p4}
\tilde{\varrho}_{t}=e^{\mathcal{L}t}\rho_{0}+e^{\mathcal{L}t}\int_{0}^{t}dse^{-\mathcal{L}s}\left([\tilde{\varrho}^{01}_{s},L^{\dagger}]\xi_{s}+[L,\tilde{\varrho}^{10}_{s}]\xi_{s}^{\ast}\right)
\end{equation}
where
\begin{equation}\label{p5}
\tilde{\varrho}^{01}_{t}=e^{\mathcal{L}t}\int_{0}^{t}dse^{-\mathcal{L}s}[L,\tilde{\varrho}^{00}_{s}]\xi_{s}^{\ast},
\end{equation}
\begin{equation}\label{p6}
\tilde{\varrho}^{00}_{t}=e^{\mathcal{L}t}\rho_{0},
\end{equation}
and $\rho_{0}$ is the initial state of $\mathcal{S}$. Clearly, by taking the derivative of (\ref{p4})-(\ref{p5}) one arrives in (\ref{p1})-(\ref{p3}). 

Note that using the conditional operators $\rho_{t|t_{m},\ldots,t_{2},t_{1}}$ one can find the whole statistics of photons of the output field. For instance, the mean time of the $m$-th counts can be determined with making use of conditional operators as
\begin{equation}
\overline{t}_{m}= \int_{0}^{+\infty}dt_{m}t_{m}\ldots\int_{0}^{t_{3}}dt_{2}
\int_{0}^{t_{2}}dt_{1}\mathrm{Tr}\rho_{t_{m}|t_{m},\ldots,t_{2},t_{1}},
\end{equation}
where $\mathrm{Tr}\rho_{t_{m}|t_{m},\ldots,t_{2},t_{1}}$ is the probability density of detecting $m>0$ photons at times $t_{1}, t_{2}, \ldots, t_{m}$ such that $0<t_{1}<t_{2}<\ldots<t_{m}$ and no other photons from $0$ to $t_{m}$. For a specified system $\mathcal{S}$ and initial conditions one can easily indicate the maximum number of certain counts in the output channel from $0$ to $+\infty$. For all $m$ less than or equal to this number we have 
\begin{equation}
\int_{0}^{+\infty}dt_{m}\ldots\int_{0}^{t_{3}}dt_{2}
\int_{0}^{t_{2}}dt_{1}\mathrm{Tr}\rho_{t_{m}|t_{m},\ldots,t_{2},t_{1}}=1.
\end{equation}

\section{Conditional and unconditional evolution of a two-level atom}

Let us consider the case when the system $\mathcal{S}$ is a two-level atom with the eigenstates $|g\rangle$ and $|e\rangle$, and
\begin{equation}\label{atom1}
H_{\mathcal{S}}=-\Delta_{0}\sigma_{z},\;\;\;L=\sqrt{\Gamma}\sigma_{-},
\end{equation} 
where $\Gamma \in \mathbb{R}_{+}$, $\sigma_{-}=|g\rangle \langle e|$, $\sigma_{z}=|e\rangle \langle e|-|g\rangle \langle g|$, and $\Delta_{0}=(\omega_{c}-\omega_{0})/2$ with $\omega_{c}$ being the carrier frequency of the input wave packet. 

\subsection{Quantum trajectories and statistics of the output photons}

By inserting (\ref{atom1}) into the formulas (80) and (82) from \cite{DSC17}, we obtain  the conditional vectors referring to detection of zero photons from $0$ up to $t$ of the form 
\begin{eqnarray}
|\alpha_{t|0}\rangle &=&
\left(e^{-i\Delta_{0}t}|g\rangle\langle g| + e^{\left(i\Delta_{0}-\frac{\Gamma}{2}\right)t} |e\rangle\langle e| \right) | \psi_0 \rangle \label{cond1},\label{alpha0}\\
|\beta_{t|0}\rangle &=&
- \sqrt{\Gamma}e^{\left(i\Delta_{0}-\frac{\Gamma}{2}\right)t}\int_{0}^{t} ds\xi_{s}e^{\left(-2i\Delta_{0}+\frac{\Gamma}{2}\right)s}  |e\rangle\langle g| \psi_0 \rangle.\label{beta0}
\end{eqnarray} 
We can easily give a physical interpretation of this result. It is seen from (\ref{alpha0}) and (\ref{beta0}) that when the two-level atom was initially in the ground state and we have not observed any photons up to the time $t$, it means that the atom has not met the external photon yet and it is still in the ground state or it has already met the photon, it absorbed this photon and after the absorption the system stayed in the excited state up to $t$. If the atom was initially in the excited state and we have not observed any photons up to $t$, it implies that only one scenario has made real---the system has not met the external photon yet and it is still in the excited state (only the vector $|\alpha_{t|0}\rangle$ gives non-zero contribution to the conditional state of the system). When the conditional vectors (\ref{alpha0}) and (\ref{beta0}) are inserted
in (\ref{posteriori4}), we get the unnormalized conditional state of the two-level atom for not observing any photon up to the time $t$,
\begin{eqnarray}\label{cond0}
\rho_{t|0}&=&\left(e^{-i\Delta_{0}t}|g\rangle\langle g|+e^{\left(i\Delta_{0}-\frac{\Gamma}{2}\right)t}|e\rangle\langle e|\right)|\psi_{0}\rangle\langle\psi_{0}|\nonumber\\
&\times&\left(e^{i\Delta_{0}t}|g\rangle\langle g|+e^{\left(-i\Delta_{0}-\frac{\Gamma}{2}\right)t}|e\rangle\langle e|\right)\int_{t}^{+\infty} ds|\xi_{s}|^2\nonumber\\
&+&\Gamma e^{-\Gamma t}\left|\int_{0}^{t}ds \xi_{s}e^{\left(-2i\Delta_{0}+\frac{\Gamma}{2}\right)s}\right|^2|\langle\psi_{0}|g\rangle|^2 |e\rangle\langle e|.
\end{eqnarray}
By taking the trace of (\ref{cond0}), we obtain the probability of not detecting any photons up to $t$,
\begin{eqnarray}
P_{t}(0)&=&\left(|\langle\psi_{0}|g\rangle|^2+e^{-\Gamma t}|\langle\psi_{0}|e\rangle|^2\right)
\int_{t}^{+\infty}ds |\xi_{s}|^2\nonumber\\
&+&\Gamma e^{-\Gamma t} |\langle\psi_{0}|g\rangle|^2\left|\int_{0}^{t}ds\xi_{s} e^{\left(-2i\Delta_{0}+\frac{\Gamma}{2}\right)s}\right|^2.
\end{eqnarray}
Of course, one can easily generalize this expression to the case of an arbitrary initial state of the system. Notice that the expression in the first line refers to the possibility that the system will meet the photon after the time $t$, the term in the second line is the probability of absorption of the photon before $t$ and staying after this absorption in the excited state up to $t$. We see that $P_{t=0}(0)=1$ and, by the presence of the decay terms, we have $\lim_{t\to +\infty}P_{t}(0)=0$. 

According to (84) and (85) in \cite{DSC17}, for a detection of one photon in the interval $[t_{1},t_{1}+dt)$ and no other photons from $0$ to $t$, we obtain
\begin{eqnarray}\label{alpha1}
|\alpha_{t|t_1}\rangle &=&
\sqrt{\Gamma}e^{-i\Delta_{0}t}e^{\left(2i\Delta_{0}-\frac{\Gamma}{2}\right)t_{1}}|g\rangle\langle e | \psi_0 \rangle,
\end{eqnarray}
\begin{eqnarray}\label{beta1}
|\beta_{t|t_1}\rangle = e^{-i\Delta_{0}t}\left[ \left( \xi_{t_{1}} - \Gamma 
\int_{0}^{t_{1}}ds \xi_{s}e^{\left(2i\Delta_{0}-\frac{\Gamma}{2}\right)(t_{1}-s)} \right)  
|g\rangle\langle g| \right.\nonumber\\\left.+ e^{\left(2i\Delta_{0}-\frac{\Gamma}{2}\right)t}
\left( \xi_{t_{1}} - \Gamma 
\int_{t_{1}}^{t} ds\xi_{s}  e^{\left(2i\Delta_{0}-\frac{\Gamma}{2}\right)(t_{1}-s)} \right)|e\rangle\langle e|\right]|\psi_{0}\rangle.
\end{eqnarray}
Thus when the two-level atom was initially prepared in the ground state, then we have $|\alpha_{t|t_{1}}\rangle=0$ and the only two scenarios of events are possible. Namely, we detected the photon coming from the external field or the atom absorbed the photon before time $t_{1}$, then emitted it at interval $[t_{1},t_{1}+dt)$, and stayed in the ground state up to $t$. 
These two scenarios are described respectively by the first and second terms of the formula for $|\beta_{t|t_1}\rangle$.
If the atom was initially in the excited state, it might not meet the photon before $t$, and we observed the photon emitted by the system or the atom has met the external photon before $t$, then we detected it directly from the field or the atom emitted the photon at $t_{1}$, then absorbed the photon from the field, and after this stayed in the excited state up to $t$. 

By (88) and (89) in \cite{DSC17}, for detection of two photons at the intervals $[t_{1},t_{1}+dt)$ and $[t_{2},t_{2}+dt)$ such that $0<t_1 <t_2 <t$ and no other photons from $0$ to $t$, we have 
\begin{equation}\label{alpha2}
 |\alpha_{t|t_{2},t_1 }\rangle = 0,
 \end{equation}
 \begin{eqnarray}\label{beta2}
 |\beta_{t|t_{2},t_1}\rangle &=&
 \sqrt{\Gamma}e^{-i\Delta_{0}t}
e^{\left(2i\Delta_{0}-\frac{\Gamma}{2}\right)(t_{1}+t_{2})}\Bigg( \xi_{t_{1}}e^{-\left(2i\Delta_{0}-\frac{\Gamma}{2}\right)t_{1}}+\xi_{t_{2}}e^{-\left(2i\Delta_{0}-\frac{\Gamma}{2}\right)t_{2}}\nonumber\\ 
&-&\Gamma 
\int_{t_{1}}^{t_{2}}ds\xi_{s} e^{\left(-2i\Delta_{0}+\frac{\Gamma}{2}\right)s}
\Bigg) |g\rangle\langle e| \psi_{0} \rangle.
\end{eqnarray}
Thus if we observed two photons we are certain that the system has already met the photon ($|\alpha_{t|t_{2},t_1}\rangle = 0$). The terms in (\ref{beta2}) correspond respectively to the following scenarios: 
\begin{itemize}
\item the first photon came directly from the field and the second one was emitted by the atom, 
\item the first photon was emitted by the atom and the second one came from the field, 
\item the first photon was emitted by the atom, then the atom absorbed the photon from the field, and it emitted the photon at $t_{2}$.    
\end{itemize}
All these possibilities we have to consider when the initial value of the probability of being in the excited state is non-zero. Of course, all the others conditional vectors vanish according to the fact that in our scheme we can not detect more than two photons.    

Let us notice that the mean time of the first count is given by the formula
\begin{equation}
\overline{t}_{1}=\int_{0}^{+\infty}dt_{1}t_{1}p(t_{1}),
\end{equation}
where
\begin{eqnarray}
p(t_{1})&=&e^{-\Gamma t_{1}}\left(\Gamma\int_{t_{1}}^{+\infty}ds|\xi_{s}|^2+|\xi_{t_{1}}|^2 \right)\langle e|\rho_{0}|e\rangle\nonumber\\
&& +\left|\xi_{t_{1}}-\Gamma\int_{0}^{t_{1}}ds\xi_{s}e^{\left(2i\Delta_{0}-\frac{\Gamma}{2}\right)\left(t_{1}-s\right)}\right|^2\langle g|\rho_{0}|g\rangle,
\end{eqnarray}
where $\rho_{0}$ is the initial state of the atom. One can check that 
\begin{equation}
\int_{0}^{+\infty}dt_{1}p(t_{1})=1. 
\end{equation} 
For the mean time of the second count we have the formula 
\begin{equation}
\overline{t}_{2}=\int_{0}^{+\infty}dt_{2}t_{2}\int_{0}^{t_{2}}dt_{1}p(t_{2},t_{1})
\end{equation}
with
\begin{eqnarray}
p(t_{2},t_{1})=
\Gamma
e^{-\Gamma(t_{1}+t_{2})}\left| \xi_{t_{1}}e^{-\left(2i\Delta_{0}-\frac{\Gamma}{2}\right)t_{1}}+\xi_{t_{2}}e^{-\left(2i\Delta_{0}-\frac{\Gamma}{2}\right)t_{2}}-\Gamma 
\int_{t_{1}}^{t_{2}}ds \xi_{s} e^{\left(-2i\Delta_{0}+\frac{\Gamma}{2}\right)s}
\right|^2 \langle e| \rho_{0}|e \rangle.
\end{eqnarray} 
Of course, we can apply it only if $\langle e| \rho_{0}|e \rangle=1$, then we have
 \begin{equation}
 \int_{0}^{+\infty}dt_{2}\int_{0}^{t_{2}}dt_{1}p(t_{2},t_{1})=1.
 \end{equation} 
 Note that the probability density of the time distance between two successive clicks depends on the time of the first click which results from the time dependence of $\xi_{t}$. The second photon can be detected just after the first detection when the first photon was emitted by the system and second one came from the field (or vice versa).

\subsection{The a priori evolution}

We determine the {\it a priori} state of the system by using the conditional counting representation, namely
\begin{eqnarray}
\tilde{\varrho}_{t}=\rho_{t|0}+\int_{0}^{t}dt_{1}\rho_{t|t_{1}}+\int_{0}^{t}dt_{2}\int_{0}^{t_{2}}dt_{1}
\rho_{t|t_{2},t_{1}}.
\end{eqnarray}
By referring to (\ref{alpha0}), (\ref{beta0}), (\ref{alpha1})-(\ref{beta2}) we can find the conditional operators: $\rho_{t|0}$, $\rho_{t|t_{1}}$, $\rho_{t|t_{2},t_{1}}$, and it allows us finally to obtain the explicit expression for the {\it a priori} state of the form
\begin{eqnarray}\label{stateapriori}
\tilde{\varrho}_{t}&=&\Bigg[1-\Gamma e^{-\Gamma t}\left|\int_{0}^{t}ds \xi_{s}e^{\gamma s}\right|^2\nonumber\\&&-\langle e|\rho_{0}|e\rangle e^{-\Gamma t} \left(1-	4\Gamma \mathrm{Re}\int_{0}^{t}dt_{1}\xi_{t_{1}}^{\ast}e^{\gamma^{\ast}t_{1}}\int_{0}^{t_{1}}ds\xi_{s}e^{-\gamma^{\ast}s}\right)\Bigg]|g\rangle\langle g|\nonumber\\
&&+\langle e|\rho_{0}|g\rangle e^{-\gamma t}\left(1-2\Gamma \int_{0}^{t}dt_{1}\xi_{t_{1}}e^{-\gamma^{\ast}t_{1}}\int_{0}^{t_1}ds\xi_{s}^{\ast}e^{\gamma^{\ast}s}\right)|e\rangle\langle g|\nonumber\\
&&+\langle g|\rho_{0}|e\rangle e^{-\gamma^{\ast} t}\left(1-2\Gamma \int_{0}^{t}dt_{1}\xi_{t_{1}}^{\ast}e^{-\gamma t_{1}}\int_{0}^{t_1}ds\xi_{s}e^{\gamma s}\right)|g\rangle\langle e|\nonumber\\
&&+\Bigg[\Gamma e^{-\Gamma t}\left|\int_{0}^{t}ds \xi_{s}e^{\gamma s}\right|^2\nonumber\\&&+
\langle e|\rho_{0}|e\rangle e^{-\Gamma t}\left(1-4\Gamma \mathrm{Re}\int_{0}^{t}dt_{1}\xi_{t_{1}}^{\ast}e^{\gamma^{\ast}t_{1}}\int_{0}^{t_{1}}ds\xi_{s}e^{-\gamma^{\ast}s}\right)\Bigg]|e\rangle\langle e|,
\end{eqnarray}
where $\gamma=-2i\Delta_{0}+\frac{\Gamma}{2}$. To obtain (\ref{stateapriori}) we used Eqs. (\ref{A1})-(\ref{A3}). 
This is the general formula for the state of the two-level atom interacting with the single-photon wavepacket. We see that it depends on the initial state of the system and on the shape of the wavepackage. For the atom being initially in the ground state, we can get from (\ref{stateapriori}) the probability of the excitation of the system
\begin{eqnarray}
P(t)= \Gamma e^{-\Gamma t}\left|\int_{0}^{t}ds \xi_{s}e^{\gamma s}\right|^2,
\end{eqnarray}
which was analyzed in \cite{WMHS12,RB17}. As expected, in the long time limit $t\to +\infty$, the atom goes to the ground state.

\subsection{POVM}

Note that with the stochastic counting process, $n(t)$, we can associate for a fixed time $t$ POVM $\{M_{t|m}\}$   labeled by the number of counts $m=0,1,2$. One can check that $3$-element POVM has here the form
	\begin{eqnarray}\label{POVMa}
	M_{t|0} &=& \left( \int_{t}^{+\infty} ds|\xi_{s}|^2  + \Gamma e^{-\Gamma t} \left| \int_{0}^{t} ds \xi_{s}e^{\gamma s} \right|^2 \right) | g \rangle \langle g | + e^{-\Gamma t} \int_{t}^{+\infty} ds|\xi_{s}|^2 | e \rangle \langle e |, \\
	M_{t|1} &=& \int_{0}^t dt_{1}\left| \xi_{t_{1}} -\Gamma \int_{0}^{t_1} ds\xi_{s} e^{-\gamma(t_{1} -s)} \right|^2 | g \rangle \langle g | \nonumber\\
	&&+ \Bigg[\left( 1-e^{-\Gamma t}\right) \int_{t}^{\infty} |\xi_{s}|^2 ds
	+ e^{-\Gamma t} \int_{0}^{t}dt_{1}\left| \xi_{t_{1}} -\Gamma \int_{t_1}^{t} ds\xi_{s} e^{-\gamma(t_{1} -s)} \right|^2\Bigg]
	| e \rangle \langle e |,\label{POVMb} \\
	M_{t|2} &=& \Gamma\int_{0}^{t}dt_{2} \int_{0}^{t_2}dt_{1}  e^{-\Gamma(t_1+t_2)}\left| \xi_{t_1} e^{\gamma t_{1}} + \xi_{t_2} e^{\gamma t_{2}} 
	-\Gamma\int_{t_1}^{t_2} ds\xi_{s} e^{\gamma s} \right|^2 | e \rangle \langle e |.\label{POVMc}
	\end{eqnarray}
 Clearly, for any initial state of the two-level atom,  $\rho_{0}$, $P_{t}(m)=\mathrm{Tr}M_{t|m}\rho_{0}$ is the probability of $m$ detections from $0$ to $t$. A detailed discussion on $\{M_{t|m}\}$ one can find in Appendix. By applying the POVM we can determine the statistical moments 
 \begin{equation}
 \overline{m^{k}_{t}}=\sum_{m=0}^{2}P_{t}(m)m^{k}
\end{equation}
referring to the output field. For the first moment, $\overline{m}_{t}$, being the mean number of counts in the period from $0$ to $t$, we obtain the formula
	\begin{eqnarray}\label{mom1}
\overline{m_{t}}&=&\Bigg(\int_{0}^t ds\left| \xi_{s}\right|^2-\Gamma e^{-\Gamma t} \left| \int_{0}^{t} ds \xi_{s}e^{\gamma s} \right|^2 \Bigg)
\langle g |\rho_{0}|g \rangle\nonumber\\
&& +\Bigg(\left( 1-e^{-\Gamma t}\right) \left(1+\int_{0}^{t} ds|\xi_{s}|^2\right)+ e^{-\Gamma t} \int_{0}^{t}ds|\xi_{s}|^2 -\Gamma e^{-\Gamma t}\left|\int_{0}^{t}ds \xi_{s}e^{\gamma s}\right|^2\nonumber\\
&&+	4\Gamma e^{-\Gamma t}\mathrm{Re}\int_{0}^{t}dt_{1}\xi_{t_{1}}^{\ast}e^{\gamma^{\ast}t_{1}}\int_{0}^{t_{1}}ds\xi_{s}e^{-\gamma^{\ast}s}\Bigg)
\langle e |\rho_{0}|e \rangle.
\end{eqnarray}
In the long time limit we get simply $\lim_{t\to +\infty}\overline{m_{t}}=1+\langle e |\rho_{0}|e \rangle$. 
And for the second moment, we have
	\begin{eqnarray}\label{mom2}
\overline{m_{t}^2}&=&\Bigg(\int_{0}^t ds\left| \xi_{s}\right|^2-\Gamma e^{-\Gamma t} \left| \int_{0}^{t} ds \xi_{s}e^{\gamma s} \right|^2 \Bigg)
\langle g |\rho_{0}|g \rangle\nonumber\\
&& +\Bigg(\left( 1-e^{-\Gamma t}\right) \left(1+3\int_{0}^{t} ds|\xi_{s}|^2\right)
+ e^{-\Gamma t} \int_{0}^{t}ds|\xi_{s}|^2 -3\Gamma e^{-\Gamma t}\left|\int_{0}^{t}ds \xi_{s}e^{\gamma s}\right|^2\nonumber\\
&&+	12\Gamma e^{-\Gamma t}\mathrm{Re}\int_{0}^{t}dt_{1}\xi_{t_{1}}^{\ast}e^{\gamma^{\ast}t_{1}}\int_{0}^{t_{1}}ds\xi_{s}e^{-\gamma^{\ast}s}\Bigg)
\langle e |\rho_{0}|e \rangle.
\end{eqnarray}
By using (\ref{mom1}) and (\ref{mom2}) we can find the Mandel $Q_{t}$ parameter, defined as
\begin{equation}
Q_{t}=\frac{\overline{m^2_{t}}-\overline{m}^2_{t}}{\overline{m}_{t}}-1.
\end{equation}
Let us note that for the case when the two-level atom is initially in the ground state we have $Q_{t}=-\overline{m_{t}}$.

\subsection{Results for chosen pulse shapes}
As an illustration, we make now calculations for a two specific shapes of the input pulse. We consider a resonance case assuming that $\xi_{t}\in \mathbb{R}$ and $\Delta_{0}=0$. Let us first consider a square pulse defined as 
\begin{equation}
\xi_{t}=\begin{cases}
\sqrt{\frac{\Omega}{2}},&0\leq t\leq \frac{2}{\Omega}\\
0,&t>\frac{2}{\Omega}
\end{cases}.
\end{equation} 
In this case the probability of no counts up to $t$ has the form
\begin{equation}
P_{t}(0)=\begin{cases}
\left[1-\frac{\Omega}{2}t+\frac{2\Omega}{\Gamma}\left(1-e^{-\Gamma t/2}\right)^2\right]\langle g|\rho_{0}|g\rangle+e^{-\Gamma t}\left(1-\frac{\Omega}{2}t\right)\langle e|\rho_{0}|e\rangle,&0\leq t\leq \frac{2}{\Omega}\\
\frac{2\Omega}{\Gamma}e^{-\Gamma t}\left(1-e^{\Gamma /\Omega}\right)^2\langle g|\rho_{0}|g\rangle,&t>\frac{2}{\Omega}
\end{cases}.
\end{equation} 
Respectively, for a decaying exponential pulse, given as 
\begin{equation}
\xi_{t}=\sqrt{\Omega}e^{-\Omega t/2},
\end{equation} 
we obtain 
\begin{eqnarray}
P_{t}(0)&=&\left[e^{-\Omega t}+\frac{4\Omega\Gamma e^{-\Gamma t }}{\left(\Gamma-\Omega\right)^2}\left(1-e^{-(\Omega-\Gamma) t/2}\right)^2\right]\langle g|\rho_{0}|g\rangle+e^{-(\Gamma+\Omega)t}\langle e|\rho_{0}|e\rangle.
\end{eqnarray}
One can easily recognize here the roles of the parameters governing the atom excitation and the process of emission of the photon. In Figure 1 we show the mean value of photons measured in the output channel up to $t$ for the square and exponential pulses and the system being initially in the ground state for chosen values of parameters. 

\begin{figure}[h]
	\begin{center}
		\includegraphics[width=5cm]{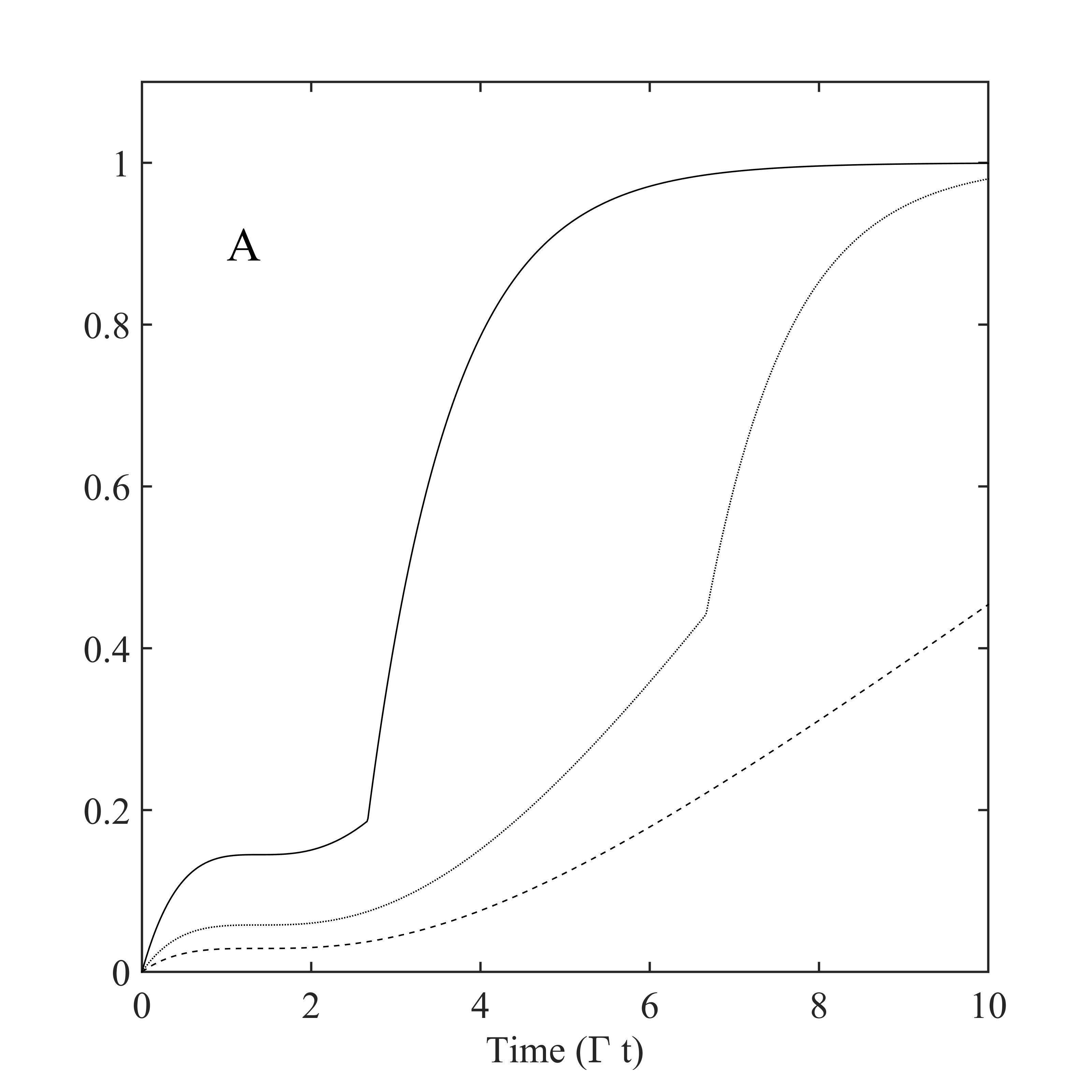}
		\includegraphics[width=5cm]{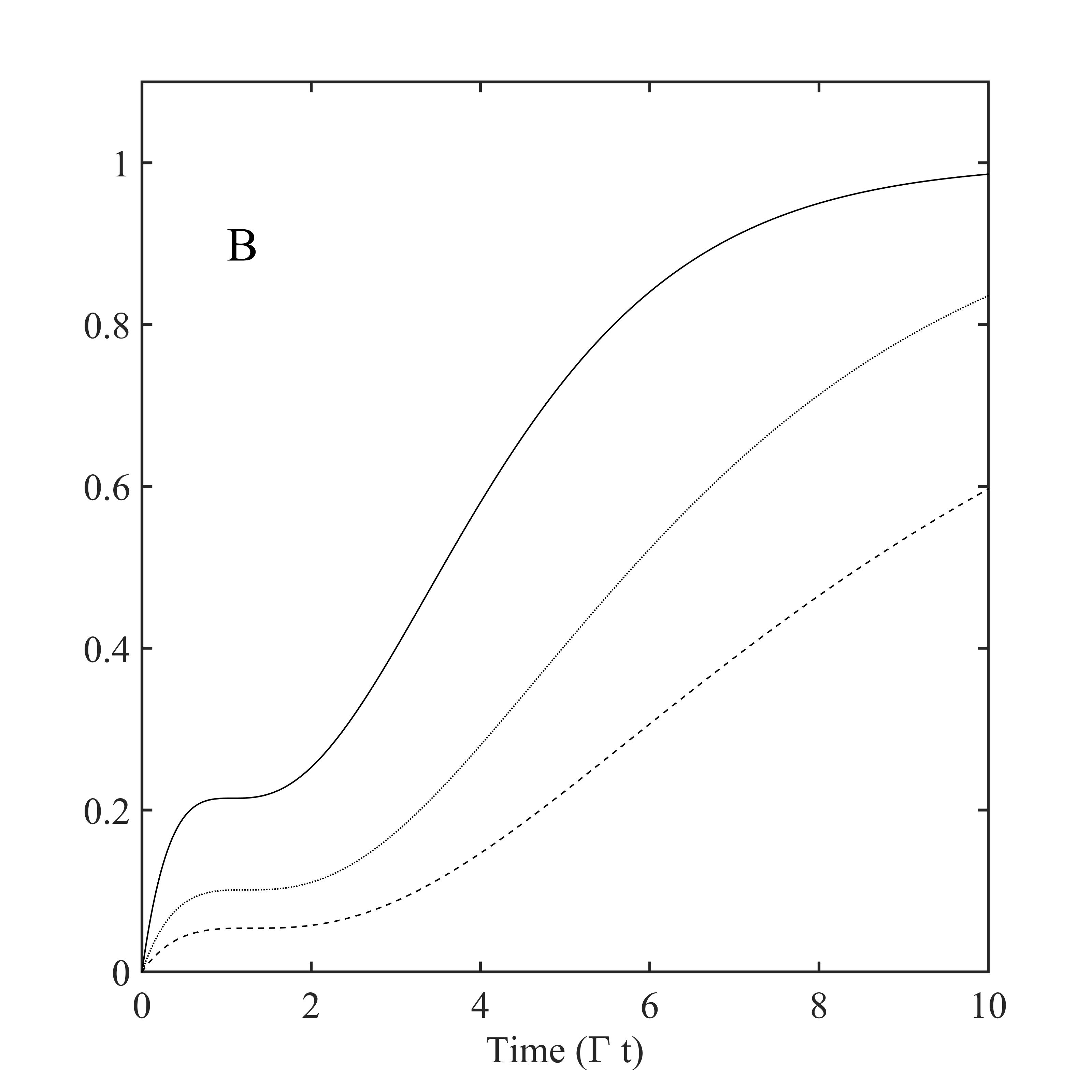}
		\caption{The mean number of photons detected from $0$ up to $t$ for the atom in the ground state for a square (A) and  exponential pulse (B) for $\Omega=0.75\Gamma$ (solid lines), $\Omega=0.30\Gamma$ (dashed lines) and $\Omega=0.15\Gamma$ (dotted lines).}
	\end{center}
\end{figure}

\section{Conclusions}

We have described the conditional and unconditional evolution of a two-level atom interacting with a single-photon pulse. In this case the system becomes correlated with the field and its stochastic evolution is given by the set of four coupled equations. In order to determine the {\it a posteriori} and {\it a priori} evolution of the two-level atom we have used the formulas for the quantum trajectories derived in \cite{DSC17}. We have expressed the states of the system in terms of the quantum trajectories related to the continuous in time measurement of photons in the output field. We have presented the formulas for the expected time of detection of the first and the second photon. Moreover, we have found the expressions for the POVM associated with the number of photons detected in the output channel up to $t$. By this we have finally determined the formulas for the mean number of photons detected from $0$ to $t$ and the Mandel $Q_{t}$ parameter defining the properties of the output signal. 

We would like to stress that the formula (\ref{apriori}) for the {\it a priori} state together with the expressions for the conditional operators given in \cite{DSC17,DSC19} is a generalization of the well known formula for the solution of the master equation \cite{Car93} to the non-Markovian case. We have used a decomposition of the quantum dynamics into quantum paths associated with detection of the photons with detector of unit quantum efficiency. This procedure allowed us not only to describe the evolution of the two-level atom but it also brought benefits by giving the analytical expressions for the probability density of photon counting. Of course, more details can be studied. Our simple model can be elaborated and experimentally tested. The experimental setup described in \cite{LSSC16} would fit perfectly to this. 

\appendix
\section{POVM}
The set (\ref{POVMa})-(\ref{POVMc}) is a set of Hermitian positive semidefinite operators that sum to the identity operator 
\begin{equation}\label{identity}
\sum_{n=0}^{2}M_{t|n}=\mathbbm{1}_{\mathcal{S}}.
\end{equation}
The first property is clearly visible, to proof that the operators sum to the identity operator, in the first step one has to 
check that 
\begin{eqnarray}\label{A1}
\int_{0}^{t}dt_{1}e^{-\Gamma t_{1}}\left|\xi_{t_{1}}e^{\gamma t_{1}}-\Gamma \int_{0}^{t_{1}}ds \xi_{s}e^{\gamma s}\right|^2
=\int_{0}^{t}ds\left|\xi_{s}\right|^2-\Gamma e^{-\Gamma t}\left|\int_{0}^{t}ds \xi_{s}e^{\gamma s}\right|^2,
\end{eqnarray}
\begin{eqnarray}\label{A2}
\int_{0}^{t}dt_{1}\left|\xi_{t_{1}}-\Gamma\int_{t_{1}}^{t}ds\xi_{s}e^{-\gamma(t_{1}-s)}\right|^2
&=&\int_{0}^{t}ds|\xi_{s}|^2+\Gamma\left|\int_{0}^{t}ds \xi_{s}e^{\gamma s}\right|^2\nonumber\\&&-4\Gamma\mathrm{Re}\int_{0}^{t}dt_{1}\xi_{t_{1}}^{\ast}e^{\gamma^{\ast}t_{1}}\int_{0}^{t_{1}}ds\xi_{s}e^{-\gamma^{\ast}s},
\end{eqnarray}
\begin{eqnarray}\label{A3}
\int_{0}^{t}dt_{2} \int_{0}^{t_2}dt_{1}  e^{-\Gamma(t_1+t_2)}
\left| \xi_{t_1} e^{\gamma t_{1}} + \xi_{t_2} e^{\gamma t_{2}} 
-\Gamma\int_{t_1}^{t_2} ds\xi_{s} e^{\gamma s} \right|^2=
\Gamma^{-1}(1-e^{-\Gamma t})\int_{0}^{t}ds|\xi_{s}|^2\nonumber\\-e^{-\Gamma t} \left|\int_{0}^{t}ds \xi_{s}e^{\gamma s}\right|^2+4e^{-\Gamma t}\mathrm{Re}\int_{0}^{t}dt_{1}\xi_{t_{1}}^{\ast}e^{\gamma^{\ast}t_{1}}\int_{0}^{t_{1}}ds\xi_{s}e^{-\gamma^{\ast}s}.
\end{eqnarray}
Making use of (\ref{A1})-(\ref{A3}), we get
	\begin{eqnarray}\label{POVM0}
	M_{t|0} &=& \left( \int_{t}^{+\infty} ds|\xi_{s}|^2  + \Gamma e^{-\Gamma t} \left| \int_{0}^{t} ds \xi_{s}e^{\gamma s} \right|^2 \right) | g \rangle \langle g | + e^{-\Gamma t} \int_{t}^{+\infty} ds|\xi_{s}|^2 | e \rangle \langle e |
	\end{eqnarray}
	\begin{eqnarray}\label{POVM1}
	M_{t|1} &=& \Bigg(\int_{0}^t ds\left| \xi_{s}\right|^2-\Gamma e^{-\Gamma t} \left| \int_{0}^{t} ds \xi_{s}e^{\gamma s} \right|^2 \Bigg)| g \rangle \langle g | \nonumber \\
	&&+ \Bigg(\left( 1-e^{-\Gamma t}\right) \int_{t}^{\infty} ds|\xi_{s}|^2
	+ e^{-\Gamma t} \int_{0}^{t}ds|\xi_{s}|^2 +\Gamma e^{-\Gamma t}\left|\int_{0}^{t}ds \xi_{s}e^{\gamma s}\right|^2,\nonumber\\
	&&-	4\Gamma e^{-\Gamma t}\mathrm{Re}\int_{0}^{t}dt_{1}\xi_{t_{1}}^{\ast}e^{\gamma^{\ast}t_{1}}\int_{0}^{t_{1}}ds\xi_{s}e^{-\gamma^{\ast}s}\Bigg)
	| e \rangle \langle e |, 
	\end{eqnarray}
	\begin{eqnarray}\label{POVM2}
	M_{t|2} &=& \Bigg(\left( 1-e^{-\Gamma t}\right) \int_{0}^{t} ds|\xi_{s}|^2
	-\Gamma e^{-\Gamma t}\left|\int_{0}^{t}ds \xi_{s}e^{\gamma s}\right|^2\nonumber\\&&	+	4\Gamma e^{-\Gamma t}\mathrm{Re}\int_{0}^{t}dt_{1}\xi_{t_{1}}^{\ast}e^{\gamma^{\ast}t_{1}}\int_{0}^{t_{1}}ds\xi_{s}e^{-\gamma^{\ast}s}\Bigg)  | e \rangle \langle e |.
	\end{eqnarray}
It is evident now that (\ref{identity}) holds.

\section*{Funding}

This paper was partially supported by the National Science Center project 2018/30/A/ST2/00837.


\begin{thebibliography}{1}
\newcommand{\enquote}[1]{``#1''}



	 \bibitem{RNB05} M.~G. Raymer, J.~Noh, K.~Banaszek, I.~A. Walmsley, ``Pure-state single-photon wavepacket generation by parametric down-conversion in a distributed microcavity," Phys. Rev. A {\bf 72} (2), 023825 (2005).

	
	\bibitem{CWSS13} M.~Cooper, L.~J. Wright, C.~S\"{o}ller, and B.~J. Smith, ``Experimental generation of multi-photon
	Fock states," Opt. Express {\bf 21}, 5309 (2013).
	
	 \bibitem{YMM13} M.~Yukawa, K.~Miyata, T.~Mizuta, H.~Yonezawa, P.~Marek, R.~Filip, A.~Furusawa, ``Generating superposition of up-to three photons for continuous variable quantum
	information processing," Opt. Express {\bf 21} (5), 5529–5535 (2013).
	
	\bibitem{PSZ13} B.~Peaudecerf, C.~Sayrin, X.~Zhou, T.~Rybarczyk, S.~Gleyzes, I.~Dotsenko, J.~M. Raimond, M.~Brune, and S.~Haroche, ``Quantum feedback experiments stabilizing Fock states of light in a cavity," Phys. Rev. A {\bf 87}, 042320 (2013).
	
	\bibitem{RR15} A.~Reiserer, G.~Rempe, ``Cavity-based quantum networks with single atoms and optical photons," Rev. Mod. Phys. {\bf 87}, 1379 (2015).
	
	 \bibitem{LMS15} P.~Lodahl, S.~Mahmoodian, and S.~Stobbe, ``Interfacing single photons and single quantum
	dots with photonic nanostructures," Rev. Mod. Phys. {\bf 87} (2), 347 (2015).
	
	\bibitem{OOM16} H.~Ogawa, H.~Ohdan, K.~Miyata, M.~Taguchi, K.~Makino, H.~Yonezawa, J.~I. Yoshikawa, and A.~Furusawa, ``Real-time quadrature measurement of a single-photon wave packet with
	continuous temporal-mode matching," Phys. Rev. Lett. {\bf 116}, 233602 (2016).
	
	\bibitem{LMS17} P.~Lodahl, S.~Mahmoodian, S.~Stobbe, A.~Rauschenbeutel, P.~Schneeweiss, J.~Volz, H.~Pichler, and P.~Zoller, ``Chiral quantum optics," Nature {\bf 541} (7638), 473 (2017).
	
	 \bibitem{SKL18} S.~Sun, H.~Kim, Z.~Luo, G.~S. Solomon, E.~Waks, ``A single-photon switch and transistor enabled by a solid-state quantum memory," Science {\bf 361} (6397), 57 (2018).
	
	
\bibitem{KB16} W.~Konyk, and J.~Gea-Banacloche, ``Quantum multimode treatment of light scattering by an atom in a waveguide," Phys. Rev. A {\bf 93}, 063807 (2016).

\bibitem{NKM15} A.~Nysteen, P.~Tr\o st Kristensen, D.~P.~S. McCutcheon, P.~Kaer, and J.~M\o rk, ``Scattering of two photons on a quantum emitter in a one-dimensional waveguide: exact dynamics and induced correlations," New J. Phys. {\bf 17}, 023030 (2015).



\bibitem{SS09} T.~Shi, and C.~P. Sun, ``Lehmann-Symanzik-Zimmermann reduction approach to multiphoton scattering in coupled-resonator arrays," Phys. Rev. B {\bf 79}, 205111 (2009).

\bibitem{SNA17} T.~F. See, Ch.~Noh, and D.~G. Angelakis, ``Diagrammatic approach to multiphoton scattering," Phys. Rev. A {\bf 95}, 053845 (2017).

\bibitem{RS16} A.~Roulet, and V.~Scarani, ``Solving the scattering of N photons on a two-level atom without computation," New J. Phys. {\bf 18}, 09303 (2016).

	
	\bibitem{GEPZ98} M.~K. Gheri, K.~Ellinger, T.~Pellizzari, and P.~Zoller, ``Photon-Wavepackets as Flying Quantum Bits," Fortschr. Phys. {\bf 46} 4-5, 401 (1998).
	
	
	\bibitem{DHR02} P.~Domokos, P.~Horak, and H.~Ritsch, ``Quantum description of light-pulse scattering on a single atom in waveguides," Phys. Rev. A {\bf 65}, 033832 (2002).
	
	
	\bibitem{GJN11} J.~E. Gough, M.~R. James, and H.~I. Nurdin, ``Quantum master equation and filter for systems driven by fields in a single photon state," in 50th {\it IEEE Conference on Decision and Control and European Control Conference} (IEEE, 2011), pp.5570 - 5576. 
	
	
	\bibitem{WMSS11} Y.~Wang, J.~Min\'{a}\v{r}, L.~Sheridan, and V.~Scarani, ``Efficient excitation of a two-level atom by a single photon in a propagating mode," Phys. Rev. A {\bf 83}, 063842 (2011).
	
	\bibitem{WMHS12} Y.~Wang, J.~Min\'{a}\u{r}, G.~H\'{e}tet, and V.~ Scarani, ``Quantum memory with a single two-level atom in a half cavity," Phys. Rev. A {\bf 85}, 013823 (2012).
	
	\bibitem{BCBC12} B.~Q. Baragiola, R.~L. Cook, A.~M. Bra\'{n}czyk, and J.~Combes, ``$N$-photon wave packets interacting with an arbitrary quantum system," Phys. Rev. A {\bf 86}, 013811 (2012).
	
	\bibitem{RB17} H.~S. Rag, and J.~Gea-Banacloche, ``Two-level-atom excitation probability for single- and $N$-photon wave packets," Phys. Rev. A  {\bf 96}, 033817 (2017).
	
 
	
	\bibitem{GJNC12a} J.~E. Gough, M.~R. James, H.~I. Nurdin, and J.~Combes, ``Quantum filtering for systems driven by fields in single-photon states or superposition of coherent states," Phys. Rev. A {\bf 86}, 043819 (2012).
	
	\bibitem{GJN12b} J.~E. Gough, M.~R. James, and H.~I. Nurdin, ``Single photon quantum filtering using non-Markovian embeddings," Phil. Trans. R. Soc. A {\bf 370}, 5408 (2012).
	
	 \bibitem{CHJ12} A. R. R. Carvalho, M. R. Hush, and M. R. James, ``Cavity driven by a single photon: conditional dynamics and non-linear phase shift," Phys. Rev. A {\bf 86}, 023806 (2012). 	
	
	\bibitem{GJN13} J.~E. Gough, M.~R. James, and H.~I. Nurdin, ``Quantum filtering for systems driven by fields in single photon states and superposition of coherent states using non-Markovian embeddings," Quantum Inf.  Process. vol. {\bf 12} Issue 3 1469 (2013).
	
	
	
	\bibitem{GJN14} J.~E. Gough, M.~R. James, and H.~I. Nurdin, ``Quantum trajectories for a class of continuous matrix product input states," New Journal of Physics {\bf 16}, 075008 (2014).
	
	\bibitem{GZ15} J.~E. Gough, and G.~Zhang, ``Generating nonclassical quantum input field states with modulating filters," EPJ Quantum Technology {\bf 2}, 15 (2015).
	
	\bibitem{DZA15} Z.~Dong, G.~Zhang, and N.~H. Amini, ``Quantum filtering for multiple measurements driven by fields in single-photon states," in Proceedings of American Control Conference (IEEE, 2016) pp. 4754-4759. 
		
	\bibitem{SZX16} H.~T. Song, G.~F. Zhang, and Z.~R. Xi, ``Continuous-mode multi-photon filtering," SIAM Journal on Control and Optimization {\bf 54}, 1602 (2016).
	
	\bibitem{PDZ16} Y.~Pan, D.~Dong, and G.~F. Zhang, ``Exact analysis of the response of quantum systems to two photons using a QSDE approach," New. J. Phys. {\bf 18}, 033004 (2016).
	
	\bibitem{BC17} B.~Q. Baragiola, and J.~Combes, ``Quantum trajectories for propagating Fock states," Phys. Rev. A {\bf 96}, 023819 (2017).
	
	\bibitem{DSC17} A. D\k{a}browska, G. Sarbicki, and D. Chru\'sci\'nski, ``Quantum trajectories for a system interacting with environment in a single-photon state: counting and diffusive processes," Phys. Rev. A {\bf 96}, 053819 (2017).
	
	\bibitem{D18} A. D\k{a}browska, ``Quantum Filtering Equations for a System Driven by Nonclassical Fields," Open Syst. Inf. Dyn. {\bf 25}, 1850007 (2018).
		
	\bibitem{DSC19} A. D\k{a}browska, G. Sarbicki, and D. Chru\'sci\'nski, ``Quantum trajectories for a system interacting with environment in $N$-photon state," Phys. A: Math. Theor. {\bf 52}, 105303 (2019).
	
	\bibitem{D19} A. D\k{a}browska, ``Quantum trajectories for environment in superposition of coherent states," Quantum Inf. Process. {\bf 18} 224 (2019). 

    \bibitem{Z19} G.~F. Zhang, ``Single-photon coherent feedback control and filtering: a survey," arXiv preprint arXiv:1902.10961
    
    \bibitem{DZA19} Z.~Dong, G.~Zhang, and N.~H. Amini, ``Quantum filtering for a two-level atom driven by two counter-propagating photons," Quantum Information Processing, {\bf 18}, 136 (2019).
    
    


\bibitem{L00} R.~Loudon, {\em The Quantum Theory of Light} third edition (Oxford University Press, Oxford, 2000).

\bibitem{M08} G.~J. Milburn, ``Coherent control of single photon states," Eur. Phys. J. Spec. Top. {\bf 159}, 113 (2008).

	
	\bibitem{Bel89} V.~P. Belavkin, ``A continuous counting observation and posterior quantum dynamics," J.Phys. A: Math. Gen. {\bf 22}, L1109 (1989).
	
	\bibitem{Bel90} V.~P. Belavkin, ``A posterior Schr\"{o}dinger equation for continuous nondemolition measurement," J. Math. Phys. {\bf 31}, 2930 (1990).
	
	\bibitem{BB91} A.~Barchielli and V.~P. Belavkin, ``Measurements continuous in time and a posteriori states in quantum mechanics,"
	J. Phys. A: Math. Gen. {\bf 24}, 1495 (1991).
		
	\bibitem{Car93} H.~Carmichael, {\em An Open Systems Approach to Quantum Optics} (Springer-Verlag Berlin-Heidelberg, 1993).
	
	\bibitem{BP02} H.~P. Breuer and F.~Petruccione, {\em The Theory of Open Quantum Systems} (Oxford University Press, New York, 2002).
	
	\bibitem{GZ10} C.~W. Gardiner and P.~Zoller, {\em Quantum noise}
	(Springer-Verlag Berlin-Heidelberg, 2010).
	
	\bibitem{WM10} W.~M. Wiseman and G.~J. Milburn, {\em Quantum measurement and control} (Cambridge University Press, 2010).
	
	\bibitem{HP84} R.~L. Hudson and K.~R. Parthasarathy, ``Quantum Ito's formula and stochastic evolutions," Commun. Math. Phys. {\bf 93}, 301 (1984).
	
	\bibitem{Par92} K.~R. Parthasarathy, {\it An Introduction to
		Quantum Stochastic Calculus} (Basel: Birkh{\"{a}}user Verlag, 1992).

	\bibitem{GarCol85} C.~W. Gardiner and M.~J. Collet, ``Input and output in damped quantum systems: Quantum stochastic differential equations and the master equation," Phys. Rev. A  {\bf 31}, 3761 (1985).
	
	\bibitem{Bar06} A.~Barchielli, ``Continual Measurements in Quantum Mechanics and Quantum Stochastic Calculus" in {\em Lecture Notes Math.} 1882, S.~Attal, A.~Joye, and C-A. Pillet, eds. (Springer, Berlin, 2006) pp. 207-291
	
		



	
	
	\bibitem{GKS76} V.~Gorini, A.~Kossakowski, and E.~C. G. Sudarshan, ``Completely positive dynamical semigroups of $N$-level systems," J. Math. Phys. {\bf 17}, 821 (1976).
	
	\bibitem{Lin76} G.~Lindblad, ``On the generators of quantum dynamical semigroups," Comm. Math. Phys. {\bf 48}, 119 (1976).
	

	
	
	\bibitem{AP06} S.~Attal and Y.~Pautrat, ``From Repeated to Continuous Quantum Interactions," Ann. Henri Poincar\'{e} {\bf 7}, 59 (2006).
	
	\bibitem{P08} C.~Pellegrini, ``Existence, uniqueness and approximation of a stochastic Schr\"{o}dinger equation: The diffusive case," Ann. Probab. {\bf 36}, No. 6 2332 (2008).
	
	\bibitem{PP09} C.~Pellegrini and F.~Petruccione, ``Non-Markovian quantum repeated interactions and measurements," J. Phys. A Math. Theor. {\bf 42}, 425304 (2009).
	
	
	\bibitem{P10} C.~Pellegrini, ``Existence, uniqueness and approximation of the jump-type stochastic Schr\"{o}dinger equation for two-level systems," Stoch. Proc. Appl. {\bf 120} Issue 9, 1722 (2010).
	

	
	
	\bibitem{KLS16} S.~Kretschmer, K.~Luoma, and W.~T. Strunz, ``Collision
	model for non-Markovian quantum dynamics," Phys. Rev. A {\bf 94}, 012106 (2016).
	
	\bibitem{C17} F.~Ciccarello, ``Collision models in quantum optics," Quantum Measurements and Quantum Metrology {\bf 4},  53 (2017).
	
	\bibitem{ACMZ17} N.~Altamirano, P.~Corona-Ugalde, R.~B. Mann, and M. Zych, ``Unitarity, feedback, interactions-dynamics emergent from repeated measurements," New. J. Phys. {\bf 19}, 013035 (2017).
	
	\bibitem{FPMZ17} S.~N. Filippov, J.~Piilo, S.~Maniscalco, and M.~Ziman, ``Divisibility of quantum dynamical maps and collision models," Phys. Rev. A {\bf 96}, 032111 (2017).
	
	\bibitem{GCMC18} J.~A. Gross, C.~M. Caves, G.~J. Milburn, and J.~Combes, ``Qubit models of weak continuous measurements: markovian conditional and open-system dynamics," Quantum Sci. Technol. {\bf 3} 024005 (2018).
	
	
	\bibitem{CC19} D.~Cilluffo and F.~Ciccarelli, ``Quantum Non-Markovian Collision Models from Colored-Noise Baths," in Advances in Open Systems and Fundamental Tests of Quantum Mechanics. Springer Proceedings in Physics, vol 23, B.~Vacchini, H.~P. Breuer, and A.~Bassi, eds. (Springer, Cham, 2019), pp. 29-40. 
	


	\bibitem{RFZB12} T. Ryb\'{a}r, S. N. Filippov, M. Ziman, and V. Bu\u{z}ek, ``Simulation of indivisible qubit channels in collision models," J. Phys. B {\bf 45}, 154006 (2012).
	
	\bibitem{BCMS14} N. K. Bernardes, A. R. R. Carvalho, C. H. Monken, M. F. Santos, ``Environmental correlations and Markovian to non-Markovian transitions in collisional models," Phys. Rev A {\bf 90}, 032111 (2014).
	

	
	\bibitem{BCMS17} N. K. Bernardes, A. R. R. Carvalho, C. H. Monken, M. F. Santos, ``Coarse graining a non-Markovian collisional model," Phys. Rev. A {\bf 95}, 032117 (2017).
	
	

	
	
	\bibitem{MV17} E. Mascarenhas, and I. de Vega, ``Quantum critical probing and simulation of colored quantum noise", Phys. Rev. A {\bf 96}, 062117 (2017).
	

   
   
   \bibitem{B02} T. A. Brun, ``A simple model of quantum trajectories," Amer. J. Phys. {\bf 70}, 719 (2002).
   
   \bibitem{M95} P. A. Meyer, {\em Quantum Probability for Probabilists} (Springer-Verlag, Berlin Heidelberg 1995). 
   
   \bibitem{GS04} J. E. Gough, A. Sobolev, ``Stochastic Schr\"odinger Equations as Limit of Discrete Filtering," Open Syst. Inf. Dyn. {\bf 11}, 235 (2004). 
	
   \bibitem{BHJ09} L. Bouten, R. Handel, and M. R. James, ``A discrete invitation to quantum filtering and feedback control," {\it SIAM REVIEW} {\bf 51}, 239 (2009).
   
   \bibitem{FTVRS18} A. K. Fischer, R. Trivedi, V. Ramasesh, I. Siddiqi, and J. Vu\u{c}kovi\'{c}, ``Scattering into one-dimensional waveguides from a coherently-driven quantum-optical system," Quantum {\bf 2}, 69 (2018).
   
   
 \bibitem{LSSC16} V. Leong, M. A. Seidler, M. Steiner, A. Cere, and Ch. Kurtsiefer,``Time-resolved scattering of a single photon by a single atom," Nat. Commun. {\bf 7}, 13716 (2016).
   
  
 
	
\end{thebibliography}
\end{document}